\documentclass{article}
\usepackage{spconf,amsmath,graphicx}

\usepackage[utf8]{inputenc} 
\usepackage[T1]{fontenc}    
\usepackage{hyperref}       
\usepackage{url}            
\usepackage{booktabs}       
\usepackage{amsfonts}       
\usepackage{nicefrac}       
\usepackage{microtype}      
\usepackage{amsmath, amssymb, bbm, mathtools}
\usepackage{algorithm}
\usepackage[noend]{algpseudocode}
\usepackage{subcaption}
\usepackage{graphicx}
\usepackage{color}
\usepackage{cite}
\usepackage{siunitx}

\renewcommand{\vec}[1]{\boldsymbol{\mathbf{#1}}}

\newcommand{\R}{\mathbb{R}}

\newcommand{\matx}[1]{\boldsymbol{\mathrm{#1}}}

\usepackage[nomargin,inline,draft]{fixme}
\fxsetup{theme=color,mode=multiuser}
\FXRegisterAuthor{pf}{apf}{\color{red}pascal}

\makeatletter
\def\@name{\emph{Clément Vignac}\qquad \emph{Guillermo Ortiz-Jiménez}\qquad \emph{Pascal Frossard}}
\makeatother

\def\Nobs{{N_\textit{obs}}}
\title{On the choice of graph neural network architectures}
\address{École Polytechnique Fédérale de Lausanne (EPFL), Lausanne, Switzerland}
\begin{document}
%
\maketitle
\begin{abstract}
Seminal works on graph neural networks have primarily targeted semi-supervised node classification problems with few observed labels and high-dimensional signals. With the development of graph networks, this setup has become a de facto benchmark for a significant body of research. Interestingly, several works have recently shown that in this particular setting, graph neural networks do not perform much better than predefined low-pass filters followed by a linear classifier. However, when learning from little data in a high-dimensional space, it is not surprising that simple and heavily regularized methods are near-optimal. In this paper, we show empirically that in settings with fewer features and more training data, more complex graph networks significantly outperform simple models, and propose a few insights towards the proper choice of graph network architectures. We finally outline the importance of using sufficiently diverse benchmarks (including lower dimensional signals as well) when designing and studying new types of graph neural networks.

\end{abstract}


\section{Introduction}\label{sec:intro}

Machine learning on graphs has found many applications in domains such as quantum chemistry \cite{schutt2018schnet}, reinforcement learning \cite{bapst2019structured} or point cloud processing \cite{wang2019dynamic}. Recently, this research field has undergone a new technological revolution with the introduction of graph convolutional networks, which permitted to extend the success of deep learning to irregular domains\cite{bronstein2017gdl}. However, there is no single definition of a graph convolution, and different methods have been proposed for different applications. \smallskip

Convolutional networks for graphs were first introduced in the spectral domain using the graph Fourier transform \cite{henaff2015deep}. This required to perform an eigendecomposition of the graph Laplacian, which is computationally too costly for large graphs. In \cite{defferrard2016convolutional,khasanova:ICML2017}, reparameterizations of the filters by Laplacian polynomials were proposed, which allowed efficient computations and opened the way to a large variety of applications. In order to benchmark their method, \cite{defferrard2016convolutional} tackled graph classification, a task that requires graph convolutions but also graph coarsening. This significantly complicates the method, thus making it more difficult to compare different types of convolutions.

This methodological problem was simplified by the authors of \cite{kipf2016semi}, who designed a simpler version of Laplacian polynomials called graph convolutional networks (GCN), and proposed to evaluate on node classification instead\footnote{For this purpose, they used the Planetoid citation datasets, which consist of citation networks between scientific papers, where each node has a signal corresponding to a bag-of-word representation of the paper.}. Because it does not require coarsening, node classification allows for simpler architectures, and hence this became the reference task to compare graph networks. Since then, neural networks of increasing complexity have been designed and studied in a similar context. \smallskip

However, the choice of the benchmark datasets has strong implications on the development of new methods. Indeed, in the standard setting, simple graph networks have been shown to perform on par with sophisticated ones when hyperparameters are tuned fairly \cite{shchur2018pitfalls}. Hence, some works \cite{wu2019simplifying,maehara2019revisiting,klicpera2018predict} have advocated for simplifying GCNs even further, claiming that linear classification methods are powerful enough to solve this task. \smallskip

In this work, we hypothesize that the success of simplified graph neural network architectures can be explained by the statistics of the benchmark datasets. Indeed, semi-supervised learning on a graph with $N_{\textit{obs}}$ observed nodes and $D$ dimensional graph signals can be seen in a first approximation as a classification problem with $N_{\textit{obs}}$ training examples in $\mathbb R^D$. When $N_{\textit{obs}} \ll D$, it is known that heavily regularized linear models are optimal, and $D$ is typically of the order of 10 $N_\textit{obs}$ in standard benchmark datasets. By varying the number of training points or features in the dataset, we show experimentally\footnote{Our source code is available at \url{github.com/LTS4/gnn_statistics}.} that more complex models becomes more effective when the ratio $N_{obs} / D$ grows. This confirms that more diverse benchmarks are necessarily to evaluate fairly the performance of the different graph network architectures. \smallskip

Finally, we provide some insights towards the proper design of graph neural networks. Whereas increasing the complexity of the neural network is beneficial when more data is available, we find that intertwining propagation and learning is not necessary to obtain good performance. This allows to use simpler architectures: since propagation can be treated as a preprocessing step, the graph structure is not used during training, which results in a computational gain.

\section{Problem statement}\label{sec:statement}

We consider a graph $\mathcal{G}$ with $N$ nodes and write its weighted adjacency matrix $\matx{A}\in\R^{N\times N}$. We denote by $\matx{D}=\operatorname{diag}(\matx{A}\vec{1})$ the matrix containing the degree of each node in its main diagonal, $\vec{1}$ being the ones vector. We assume that each node carries a signal or feature vector $\vec{x}_i\in\R^D$ and possibly a label $y_i\in\{1,\dots,C\}$. These feature vectors are aggregated in $\matx{X}\in\R^{N\times D}$. \smallskip

The goal of node classification is the following: given $\matx{A}$, $\matx{X}$ and a small subset of labels, one tries to predict the value of the remaining labels of the graph. Node classification can be viewed as an example of semi-supervised learning \cite{chapelle2009semi}, where unlabeled data observed during training can be leveraged to achieve better performance. The recently developed Graph Neural Networks are the most popular framework to address node classification. However, there is now a plethora of different graph networks \cite{wu2019comprehensive}, and it is important to develop insights about the proper design choice for a given dataset. \smallskip

In supervised learning, the choice of the model is typically guided by complexity measures such as the VC dimension. In particular, two crucial parameters are the number of training points $N$ and the dimension of the space $D$: more complex classes of functions can be used when $N$ is large and $D$ small. In semi-supervised learning however, we currently do not have such complexity measures. As a first approximation, we can consider only the labeled nodes and view the problem as the classification of points in dimension $D$ given $\Nobs$ training examples. The ratio $\Nobs/D$ is however extremely small for common benchmark datasets (cf. Table \ref{tab:statistics}). If we now use insights from supervised learning, we would expect heavily regularized linear methods to be optimal, or close to it, for this classification problem. This has been recently verified in \cite{wu2019simplifying}. However, the very specific properties of these benchmark datasets do not permit to confidently extend these insights to other settings. In particular, we believe that the good performance of simple graph neural networks is a consequence of the current evaluation strategies focusing on the small $N_{obs}/D$ regime. Therefore, we make the following hypothesis:

\smallskip
\emph{Depending on the data structure, and especially on the ratio $N_{obs}/D$, graph neural networks of different complexities are appropriate, with no model being universally superior to all other ones.}
\smallskip

We will verify experimentally this hypothesis and show in particular that, in settings where $N_{obs}/D$ is high, non-linear models perform significantly better.

\section{Graph neural networks}\label{sec:gcn}
\begin{table}[t!]
    \centering
    \begin{tabular}{l r r r}
    \textit{Statistics}   & Pubmed & Citeseer & Cora \\ \hline
     Nodes (observed)        & $19717$ ($60$) & $2110$ ($120$) & $2485$ ($140$) \\
     Features    & $500$ & $3703$ & $1433$   \\ 
     $\Nobs / D$  & $0.11$& $0.03$ & $0.10$ \\     \addlinespace[0.3em]
     \textit{Accuracy} \\ \hline
     GCN \cite{kipf2016semi}      & $77.4\pm0.5$ & $68.1\pm0.3$ & $79.4\pm0.4$\\
     SGC \cite{wu2019simplifying} & $76.5\pm0.5$ & $68.7\pm0.3$ & $80.2\pm0.3$\\
     APPNP \cite{klicpera2018predict} & $\textbf{79.4}\pm0.4$ & $\textbf{70.0}\pm0.3$ & $\textbf{82.2}\pm0.3$
    \end{tabular}
    \caption{Statistics of the Planetoid datasets and average accuracy (with 95\% confidence interval) on 100 splits using 20 observed nodes per class (numbers from \cite{fey2019fast}). In this setting, linear methods such as SGC and APPNP are very competitive.}
    \label{tab:statistics}
\end{table}

Graph neural networks provide a general method to address node classification. They define a parameterized and differentiable function $h:\R^{N\times D}\rightarrow \R^{N \times C}$ that can efficiently be trained to minimize a relevant loss, e.g. empirical cross-entropy or mean squared error on the training data. In order to introduce some form of domain prior\footnote{Compared to other approaches, graph neural networks have the advantage of using both the the graph structure (encoded in the adjacency matrix) and the information contained in the node features.}, virtually all graph neural networks are formed using a composition of i) propagation steps $\varphi:\R^N\rightarrow\R^N$, which are applied column-wise to $\matx{X}$ using the structure of $\matx{A}$, and ii) feature extractors $f:\R^D\rightarrow \R^{D'}$, which act row-wise on $\matx{X}$. \smallskip

In this sense, most graph neural networks only differ on the way they define $\varphi$ and $f$, and the way these two are composed. We briefly review the most relevant graph neural networks for this work.

\paragraph*{Graph convolutional networks (GCN)} Without any doubt, the \emph{de facto} standard graph neural network is the GCN \cite{kipf2016semi}. It consists in the consecutive application of one-hop aggregation steps $\varphi(\matx{X})=\tilde{\matx{A}}\matx{X}$, based on the normalized adjacency matrix with self-loops $\tilde{\matx{A}}=(\matx{D}+\matx{I})^{-1/2}(\matx{A}+\matx{I})(\matx{D}+\matx{I})^{-1/2}$, and simple non-linear feature extractors $f(\matx{X})=\sigma(\matx{X}\matx{W})$, where $\matx{W}\in\R^{D\times D'}$ represents a set of coefficients to be learned and $\sigma$ is a point-wise non-linearity, e.g., Rectified Linear Unit (ReLU). These two steps are applied iteratively, leading to:
\begin{equation}
    h(\matx{X})=\operatorname{softmax}\left(\tilde{\matx{A}}\sigma\left(\dots\sigma\left(\tilde{\matx{A}}\matx{X}\matx{W}_1\right)\dots\right)\matx{W}_K\right) \label{eq:gcn}
\end{equation}
Although GCN was one of the earliest model introduced, more complex extensions of this model \cite{monti2018motifnet, monti2018dual, velickovic2018graph} did not lead to significant improvements in the classification accuracy on standard datasets. For this reason, a recent trend has been to try to simplify it instead.

\begin{figure*}[ht!]
     \centering
     \begin{subfigure}[b]{\textwidth}
         \centering
         \includegraphics[width=\textwidth]{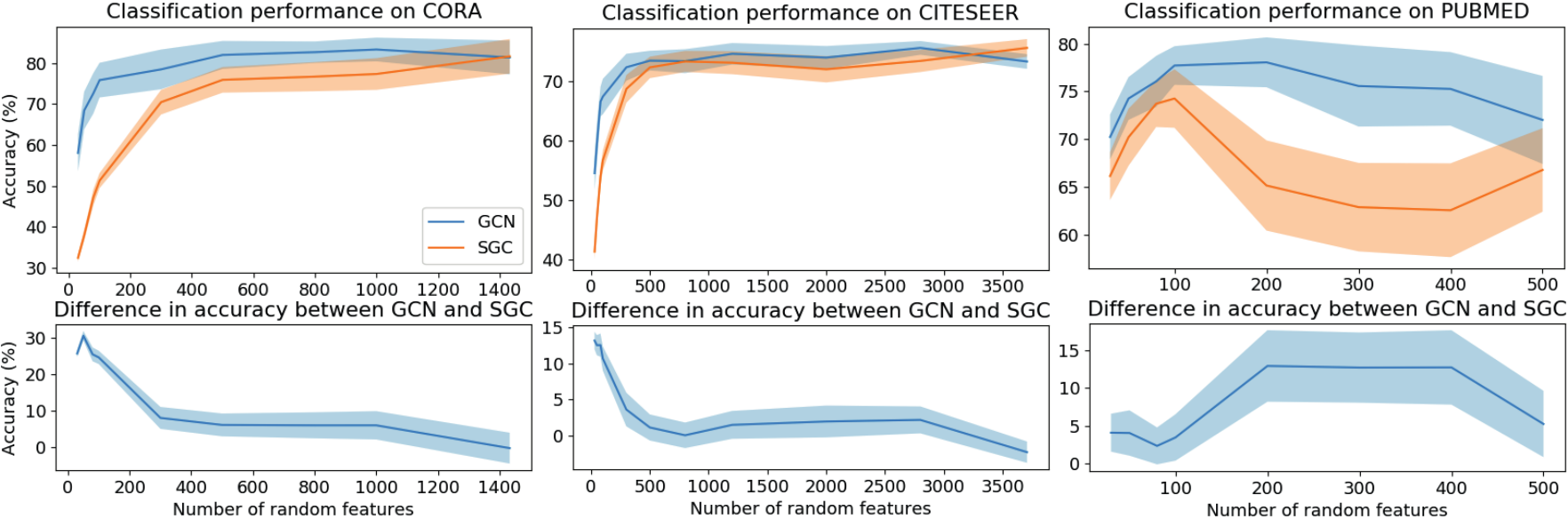}
         \caption{Varying number of random features when $50\%$ of the nodes are observed.}
         \label{fig:experiment2a}
     \end{subfigure}
     \begin{subfigure}[b]{\textwidth}
         \centering
         \includegraphics[width=\textwidth]{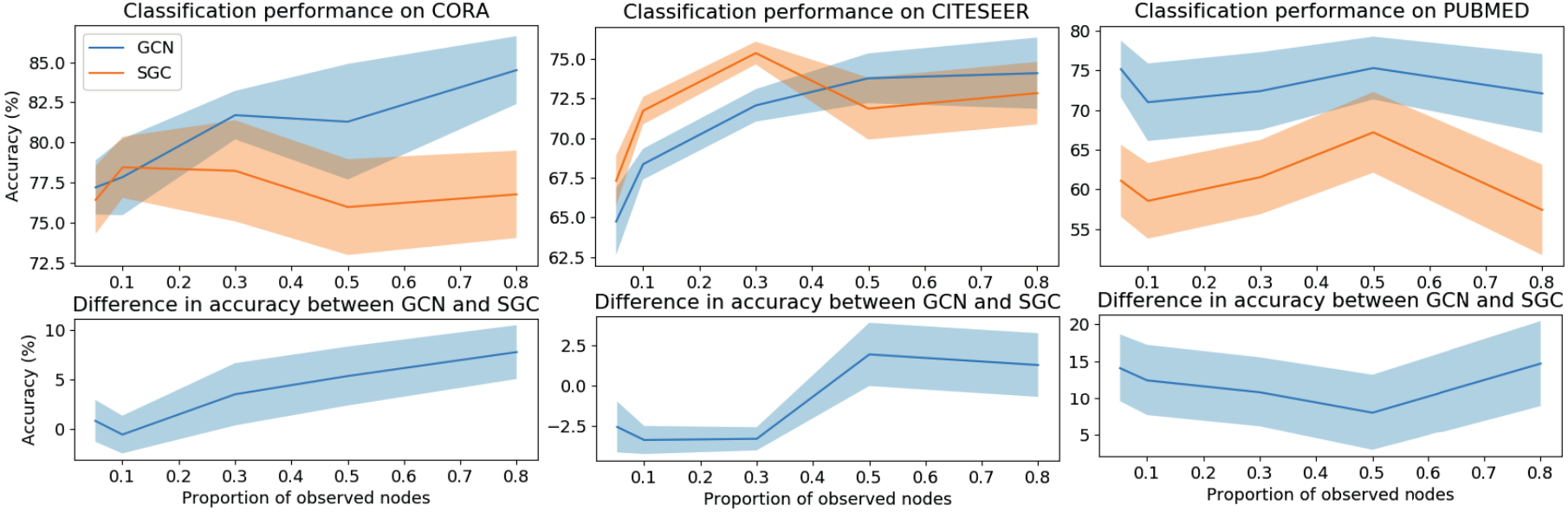}
         \caption{Varying proportion of observed nodes for 300 random features.}
         \label{fig:experiment2b}
     \end{subfigure}
\caption{Each parameter configuration was tested 40 times on random splits of the data. Shaded regions represent the $95\%$ confidence interval for the mean value.}
\vspace{-0.2em}
\end{figure*}

\paragraph*{Simplified graph convolutions (SGC)} Following this line of thought, \cite{wu2019simplifying} proposed to simplify the structure in \eqref{eq:gcn} by removing the intermediate non-linearities. By doing this, the learnable parameters $\matx W_1, ..., \matx W_k$ collapse into a single matrix $\matx W$, and the model becomes:
\begin{equation}
    h(\matx{X})= \operatorname{softmax}\left(\tilde{\matx{A}}^K\matx{X}\matx{W}\right)\label{eq:sgc}
\end{equation}
Hence, they reduced the propagation to $\varphi(\matx{X})=\tilde{\matx{A}}^K\matx{X}$, which can be interpreted as a low-pass spectral filter in the graph Fourier domain\footnote{It can be shown \cite{wu2019simplifying,ortega2018gsp} that the operation $\varphi(\matx{X})=\tilde{\matx{A}}^K\matx{X}$ is equivalent to a low-pass graph filter of the symmetric normalized Laplacian $\matx D^{-1/2} \matx A \matx D^{-1/2}$.}, and they effectively simplify the learning task to a logistic regression on these filtered features. Despite its simplicity, SGC performs on par with GCN on standard benchmarks (cf. Table \ref{tab:statistics}).

\paragraph*{Approximate personalized propagation of neural predictions (APPNP)} In theory, the number of layers $K$ of a GCN \eqref{eq:gcn} or a SGC \eqref{eq:sgc} can be arbitrary. However, high values of $K$ are rarely used in practice. The reason is that multiple applications of the adjacency operator on the features tend to produce an important smoothing effect on the signal used for learning, which harms classification accuracy \cite{oono2020graph}. Inspired by the success of the personalized page rank algorithm, the authors of \cite{klicpera2018predict} proposed to circumvent this issue by modifying the propagation strategy to $
    \varphi(\matx{X})=\alpha\left(\matx{I}-(1-\alpha)\tilde{\matx{A}}\right)^{-1}\matx{X}$.
This operation is approximated using fixed point iterations, while a simple neural network is chosen for the feature extractor $f$. State-of-the art results on standard datasets were obtained using a linear feature extractor, in which case the model writes:
\begin{equation}
    h(\matx{X})=\alpha\left(\matx{I}-(1-\alpha)\tilde{\matx{A}}\right)^{-1} \matx{X} ~\matx{W}
\end{equation}
Nevertheless, most performance reviews for graph networks have only been conducted on datasets with very similar properties. Hence, it is important that benchmarking results are considered beyond this specific context. By doing this, we will show that simple models do not perform well in all settings, and that the network complexity should be tuned depending on the training set size and number of features. Besides, we do not find any evidence confirming that propagation and learning should be intertwined.

\section{Experimental results}\label{sec:experiments}

All our experiments are based on the experimental platform introduced in \cite{shchur2018pitfalls} in which multiple graph neural networks can be tested against the standard benchmark datasets. We repeat experiments with different train-test splits (selected uniformly at random) for each configuration of parameters to avoid overfitting to a particular training scenario.\smallskip

The main hypothesis we want to test is if, depending on the ratio of observed nodes to feature dimensionality $N_{obs}/D$, the methods introduced earlier show different behaviours. We are especially interested in the scenario in which $N_{obs}/D$ is high, since this is different from the standard evaluation regime for graph neural networks. To tune $N_{obs}/D$ we control independently the proportion of observed nodes and the dimensionality of the features. To avoid any bias in the feature selection, we change the dimensionality of our feature matrix using a random sketching matrix. That is,
\begin{equation}
    \matx{X}'=\cfrac{1}{\sqrt{D'}}~\matx{X}~\matx{W}_\text{r},
\end{equation}
where $\matx{W}_\text{r}\in\R^{D\times D'}$ is a random matrix with entries drawn from a normal distribution, and $D'$ is the target dimension. We use the same set of randomly scrambled features $\matx{X}'\in\R^{N\times D'}$ as input to the different methods.

\vspace{-0.3em}
\subsection{Changing the dataset statistics}
We compare the performance of a GCN and a SGC on different ratios $N_{obs}/D$. Both networks are built using their standard hyperparameters \cite{shchur2018pitfalls} and $K=2$. We recall that in the standard setting for citation data, $N_{obs}/D$ is of the order of 0.1 and both methods perform equally well (cf. Table \ref{tab:statistics}). \smallskip

We test separately the effect of reducing $D$ (cf. Figure~\ref{fig:experiment2a}) and increasing $N_{obs}$ (cf. Figure~\ref{fig:experiment2b}). Clearly, both methods only perform on par in the small data and high-dimensional regime. When we start to increase $N_{obs}/D$, the model based on a GCN tends to perform significantly better than the SGC, while the previously reported similarity \cite{wu2019simplifying} is only retained for the original configuration. The reason for these differences is rooted in the complexity of the two classifiers. Indeed, the GCN has the potential to fit non-linear functions and the bias in this estimation decreases when we increase $N_{obs}/D$.

\vspace{-0.3em}
\subsection{Decoupling feature extraction and propagation}
Furthermore, we evaluate the benefits of using non-linear classification methods and the need to intertwine propagation and feature extraction layers. To this end, we compare the performance of the GCN and the SGC from the previous experiments. We also study the APPNP architecture, and two models (APPNP-MLP and SGC-MLP) where the feature extractor $f$ is a two-layer neural network (MLP) instead of a logistic regression. Contrary to the original APPNP model, feature propagation is treated as a preprocessing step and performed before learning, so as to make gradient backpropagation faster during training. In our configuration, $50\%$ of the nodes are observed and the nodes have $300$ random features in order to explore different settings than those of standard benchmarks. \smallskip
\begin{table}[t!]

    \centering
    \begin{tabular}{l  r r  r}
                & Pubmed & Citeseer & Cora \\ \hline
    Ratio $N_{obs}/D$ & 32.9& 3.5 & 4.1\\ \hline
     SGC        & $76.3\pm3.2$ & $68.4\pm2.5$ & $70.0\pm3.2$ \\
     SGC-MLP    & $80.2\pm1.9$ & $70.8\pm2.8$ & $79.3\pm2.9$   \\ 
     GCN        & $78.2\pm3.0$ & $71.3\pm2.6$ & $80.9\pm3.8$ \\
     APPNP      & $80.4\pm2.7$ & $\textbf{72.8}\pm2.4$ & $\textbf{82.3}\pm3.2$\\
     APPNP-MLP  & $\textbf{84.0}\pm1.2$ & $70.9\pm2.6$ & $78.8\pm2.6$\\ 
    \end{tabular}
    \caption{Accuracy and 95\% confidence interval over 100 splits with 300 features and 50\% of the nodes in the training set.}
    \label{tab:experimental_results}
    \vspace{-1em}
\end{table}

Table~\ref{tab:experimental_results} summarizes the results.
First, we find that GCN and SGC-MLP perform similarly, suggesting that there is no need to use multilayer architectures that intertwine propagation and feature extraction. Indeed, the only difference between the two methods is that $\varphi$ and $f$ are alternated in GCN on not in GCN-MLP. Furthermore, we find that non-linear classification methods (GCN, SGC-MLP) can significatively outperform linear ones (SGC) and that, in high $N_{obs}/D$ scenarios, non-linear feature extractors also boost the  accuracy of methods with a complex propagation step such as APPNP. \smallskip

Overall these results support our claim that depending on the setup, graph neural networks of different complexities perform best. Furthermore, it is clear that having the possibility to tune the complexity of $\varphi$ and $f$ independently gives rise to a rich set of classification behaviours that can be optimized to better fit the structure of the data. Our study further outlines that the choice of the benchmark dataset is really critical when comparing the performance of different graph neural networks.

\section{Conclusions}\label{sec:conclusions}

We have empirically demonstrated that the surprising good performance of simple graph neural networks reported in the recent literature is essentially driven by the characteristics of the benchmark datasets. In particular, both the high-dimensionality of the features and the scarcity of labels in the standard setup make simple methods based on feature smoothing and linear classification nearly optimal. For richer datasets (in terms of $N_{obs}/D$), complex GNN models do outperform simpler ones. However, we find no evidence that intertwining propagation and feature extraction is necessary to obtain good performance: it is sufficient to tune the complexity of the feature extractor depending on the data statistics. Overall, when designing new types of graph neural networks, it is very important to use benchmarks with lower dimensional features and more observed nodes as well, and to assess performance on data whose statistics are suited to the complexity of the proposed method. 


\bibliographystyle{IEEEbib}
\bibliography{refs}

\begin{thebibliography}{10}

\bibitem{schutt2018schnet}
Kristof~T Sch{\"u}tt, Huziel~E Sauceda, P-J Kindermans, Alexandre Tkatchenko,
  and K-R M{\"u}ller,
\newblock ``{SchNet--A deep learning architecture for molecules and
  materials},''
\newblock {\em The Journal of Chemical Physics}, vol. 148, no. 24, pp. 241722,
  2018.

\bibitem{bapst2019structured}
Victor Bapst, Alvaro Sanchez-Gonzalez, Carl Doersch, Kimberly Stachenfeld,
  Pushmeet Kohli, Peter Battaglia, and Jessica Hamrick,
\newblock ``Structured agents for physical construction,''
\newblock in {\em International Conference on Machine Learning (ICML)}, 2019,
  pp. 464--474.

\bibitem{wang2019dynamic}
Yue Wang, Yongbin Sun, Ziwei Liu, Sanjay~E Sarma, Michael~M Bronstein, and
  Justin~M Solomon,
\newblock ``Dynamic graph cnn for learning on point clouds,''
\newblock {\em ACM Transactions on Graphics (TOG)}, vol. 38, no. 5, pp. 1--12,
  2019.

\bibitem{bronstein2017gdl}
M.~M. {Bronstein}, J.~{Bruna}, Y.~{LeCun}, A.~{Szlam}, and P.~{Vandergheynst},
\newblock ``{Geometric Deep Learning: Going beyond Euclidean data},''
\newblock {\em IEEE Signal Processing Magazine}, vol. 34, no. 4, pp. 18--42,
  July 2017.

\bibitem{henaff2015deep}
Mikael Henaff, Joan Bruna, and Yann LeCun,
\newblock ``Deep convolutional networks on graph-structured data,''
\newblock {\em arXiv preprint arXiv:1506.05163}, 2015.

\bibitem{defferrard2016convolutional}
Micha{\"e}l Defferrard, Xavier Bresson, and Pierre Vandergheynst,
\newblock ``Convolutional neural networks on graphs with fast localized
  spectral filtering,''
\newblock in {\em Advances in neural information processing systems (NIPS)},
  2016, pp. 3844--3852.

\bibitem{khasanova:ICML2017}
Renata Khasanova and Pascal Frossard,
\newblock ``Graph-based isometry invariant representation learning,''
\newblock in {\em International Conference on Machine Learning (ICML)}, 2017.

\bibitem{kipf2016semi}
Thomas~N Kipf and Max Welling,
\newblock ``Semi-supervised classification with graph convolutional networks,''
\newblock in {\em International Conference on Learning Representations (ICLR)},
  2017.

\bibitem{shchur2018pitfalls}
Oleksandr Shchur, Maximilian Mumme, Aleksandar Bojchevski, and Stephan
  G{\"u}nnemann,
\newblock ``Pitfalls of graph neural network evaluation,''
\newblock {\em arXiv preprint arXiv:1811.05868}, 2018.

\bibitem{wu2019simplifying}
Felix Wu, Amauri Souza, Tianyi Zhang, Christopher Fifty, Tao Yu, and Kilian
  Weinberger,
\newblock ``Simplifying graph convolutional networks,''
\newblock in {\em International Conference on Machine Learning (ICML)}, Long
  Beach, California, USA, Jun 2019, vol.~97, pp. 6861--6871.

\bibitem{maehara2019revisiting}
Takanori Maehara,
\newblock ``Revisiting graph neural networks: All we have is low-pass
  filters,''
\newblock {\em arXiv preprint arXiv:1905.09550}, 2019.

\bibitem{klicpera2018predict}
Johannes Klicpera, Aleksandar Bojchevski, and Stephan G{\"u}nnemann,
\newblock ``Predict then propagate: Graph neural networks meet personalized
  pagerank,''
\newblock {\em arXiv preprint arXiv:1810.05997}, 2018.

\bibitem{chapelle2009semi}
Olivier Chapelle, Bernhard Scholkopf, and Alexander Zien,
\newblock ``Semi-supervised learning,''
\newblock {\em IEEE Transactions on Neural Networks}, vol. 20, no. 3, pp.
  542--542, 2009.

\bibitem{wu2019comprehensive}
Zonghan Wu, Shirui Pan, Fengwen Chen, Guodong Long, Chengqi Zhang, and Philip~S
  Yu,
\newblock ``A comprehensive survey on graph neural networks,''
\newblock {\em arXiv preprint arXiv:1901.00596}, 2019.

\bibitem{fey2019fast}
Matthias Fey and Jan~Eric Lenssen,
\newblock ``Fast graph representation learning with pytorch geometric,''
\newblock {\em arXiv preprint arXiv:1903.02428}, 2019.

\bibitem{monti2018motifnet}
Federico Monti, Karl Otness, and Michael~M Bronstein,
\newblock ``Motifnet: a motif-based graph convolutional network for directed
  graphs,''
\newblock in {\em 2018 IEEE Data Science Workshop (DSW)}. IEEE, 2018, pp.
  225--228.

\bibitem{monti2018dual}
Federico Monti, Oleksandr Shchur, Aleksandar Bojchevski, Or~Litany, Stephan
  G{\"u}nnemann, and Michael~M Bronstein,
\newblock ``Dual-primal graph convolutional networks,''
\newblock {\em arXiv preprint arXiv:1806.00770}, 2018.

\bibitem{velickovic2018graph}
Petar Veličković, Guillem Cucurull, Arantxa Casanova, Adriana Romero, Pietro
  Liò, and Yoshua Bengio,
\newblock ``Graph attention networks,''
\newblock in {\em International Conference on Learning Representations}, 2018.

\bibitem{ortega2018gsp}
A.~{Ortega}, P.~{Frossard}, J.~{Kovačević}, J.~M.~F. {Moura}, and
  P.~{Vandergheynst},
\newblock ``{Graph Signal Processing: Overview, Challenges, and
  Applications},''
\newblock {\em Proceedings of the IEEE}, vol. 106, no. 5, pp. 808--828, May
  2018.

\bibitem{oono2020graph}
Kenta Oono and Taiji Suzuki,
\newblock ``Graph neural networks exponentially lose expressive power for node
  classification,''
\newblock in {\em International Conference on Learning Representations}, 2020.

\end{thebibliography}

\end{document}